\documentclass[referee]{aa}
\usepackage{natbib}
\bibpunct{(}{)}{;}{a}{}{,}

\input psfig.sty

\begin{document}

\title{The VMC Survey - XII. Star cluster candidates in
 the Large Magellanic Cloud\thanks{Based on observations made with VISTA at the Paranal Observatory 
under programme ID 179.B-2003.}}

\author{Andr\'es E. Piatti\inst{1,2}
\and
Roald Guandalini\inst{3} 
\and 
Valentin D. Ivanov\inst{4}
\and
Stefano Rubele\inst{5}
\and 
Maria-Rosa L. Cioni\inst{6,7}
\and
Richard de Grijs\inst{8,9}
\and
Bi-Qing For\inst{10}
\and
Gisella Clementini\inst{11}
\and
Vincenzo Ripepi\inst{12}
\and
Peter Anders\inst{13}
\and
Joana M. Oliveira\inst{14}
}

\institute{Observatorio Astron\'omico, Universidad Nacional de C\'ordoba, Laprida 854, 5000, C\'ordoba, Argentina;\\
\email{andres@oac.uncor.edu}
\and
Consejo Nacional de Investigaciones Cient\'{\i}ficas y T\'ecnicas, Av. Rivadavia 1917, C1033AAJ,
Buenos Aires, Argentina 
\and
Instituut voor Sterrenkunde, Celestijnenlaan 200 D BUS 2401, 3001 Heverlee, Belgium
\and
European Southern Observatory, Av. Alonso de C\'{o}rdoba 3107, Casilla 19, Santiago, Chile
\and
INAF, Osservatorio Astronomico di Padova, vicolo dell' Osservatorio 5, I-35122 Padova, Italy
\and
University of Hertfordshire, Physics Astronomy and Mathematics, College Lane, Hatfeild AL10 9AB, United Kingdom
\and
Leibniz-Institut f\"{u}r Astrophysik Potsdam, An der Sternwarte 16, 14482 Potsdam, Germany
\and
Kavli Institute of Astronomy and Astrophysics, Peking University, Yi He Yuan Lu 5, Hai Dian District, Beijing 100871, China
\and
Department of Astronomy, Peking University, Yi He Yuan Lu 5, Hai Dian District, Beijing 100871, China
\and
ICRAR, University of Western Australia, 35 Stirling Hwy, Crawley, WA 6009, Australia
\and
INAF, Osservatorio Astronomico di Bologna, via Ranzani 1, 40127 Bologna, Italy
\and
INAF, Osservatorio Astronomico id Capodimonte, via Moiariello 16, 80131 Napoli, Italy
\and
National Astronomical Observatories, Chinese Academy of Sciences, 20A Datun Road, Chaoyang District, Beijing 100012, China
\and
Lennard-Jones Laboratories, School of Physical and Geographical Sciences, Keele University, ST5 5BG, United Kingdom
}

\date{Received / Accepted}

\abstract{In this work we analyse Colour-Magnitude Diagrams (CMDs) 
 of catalogued star clusters located in the 
Large Magellanic Cloud (LMC), from a $YJK_s$ photometric data set obtained by the
Visible and Infrared Survey Telescope for Astronomy (VISTA) survey of the Magellanic Clouds system (VMC).}
{We studied a total of 98 objects of small angular size, typically $\sim$ 11.6 pc
in diameter projected towards both uncrowded tile LMC 8\_8 and crowded tile LMC 5\_5. 
They populate relatively crowded LMC fields with significant
 fluctuations in the stellar density, the luminosity function, and the colour distribution as well as
uncrowded fields.
This cluster sample is aimed at actually probing our performance in reaching the
CMD features of clusters with different ages in crowded/uncrowded fields. 
}
{We applied a subtraction procedure to statistically clean the cluster CMDs from field star contamination.
We then matched theoretical isochrones to the background-subtracted CMDs to determine
the ages and metallicities of the clusters. 
}
{
We estimated the ages 
of 65 clusters, which resulted to be in the age range 7.3 $<$ log($t$/yr) $<$ 9.55.
We also classified as chance grouping of stars 19 previoulsy catalogued clusters, two
possible cluster-like asterisms, and one unresolved cluster.
For other 8 objects, we could not find a clear star concentration in the $K_s$ images either, so 
we quoted them as cluster-like asterisms. Finally, we found
two previously catalogued single star clusters to be unresolved background
galaxies (KMHK747, OGLE366), and one to be a triple cluster system (BSDL 2144).
}
{}

\keywords{
techniques: photometric -- galaxies: individual: LMC -- Magellanic 
Clouds -- galaxies: star clusters.}

\titlerunning{LMC star clusters}
\authorrunning{A.E. Piatti et al.}

\maketitle

\markboth{A.E. Piatti et al.: LMC star clusters }{}

\section{Introduction}

The Magellanic Clouds are the nearest example of interacting dwarf irregular galaxies. Because of their
 distance (50-60 kpc) we can resolve individual stars in the field population and in star clusters. 
Compared to the Milky Way the Magellanic Clouds have a lower metallicity and host star clusters spanning 
the entire age range \citep{dga06,dgg08}. The Magellanic Clouds contain a few thousands star clusters 
(Bica et al. 2008, 
hereinafter B08) and represent an important laboratory for studies of stellar evolution. The sample of
 star clusters with measurements of size, mass and other parameters is, however, modest and corresponds 
to less than half the number of candidate star clusters  (e.g., Hill \& Zaritsky 2006, Werchan \& 
Zaritsky 2011, Glatt et al. 2010, Baumgardt et al. 2013, Piatti 2014).

Taking advantage of the high sensitivity and spatial resolution of the VISTA near-infrared $YJK_s$
 survey of the Magellanic Clouds system (VMC; Cioni et al. 2011) we embarked on an homogeneous 
determination of star cluster parameters. Compared to the wide-scale Magellanic Clouds Photometric 
Survey data 
(Zaritsky et al. 2002, 2004) the VMC survey data corresponds to an improvement of about a factor of 
two in pixel scale and seeing. In addition, the VMC makes use of the near-infrared filters, $YJK_s$, covers 
a wider area around each Cloud, and includes the Magellanic Bridge. The VMC covers $\sim 
170$ deg$^2$ of the entire Magellanic system with $110$ individual tiles; each tile covers $\sim1.5$ deg$^2$. 
With a statistical sample of characterised star clusters as complete as possible we will be able to 
answer some key open questions in star cluster studies, such as: Has the field experienced the same star
 formation history as the cluster stellar population? What is the distribution of star clusters as a 
function of age and metallicity? What galaxy structure is defined by star clusters with different ages?
 Is there a relation between the age and size of star clusters?

This is the first VMC paper that provides information on the star clusters of the Magellanic system. 
Some preliminary results from the analysis of star clusters in a tile covering the South Ecliptic Pole 
(SEP) region are published in Cioni et al. (2011). The complete study of star clusters in the SEP tile 
(tile LMC 8$\_$8) is presented here along with analysis of clusters in tile LMC 5$\_$5 covering the
LMC bar. Study of star clusters in other tiles will follow.
We plan to study known star clusters identified in previous 
studies, and included in the B08's catalogue, and to 
search for new star clusters based on the stellar surface density 
method (Ivanov et al. 2002; Borissova et al. 2003). Fig. 1 depicts the spatial distribution of
B08's catalog of star clusters,  wherein black points and green circles represent the
B08's catalogued star clusters and those with age estimates
available. The VMC tile distribution is superimposed.

This paper is organised as follows. VMC observations and data reduction are presented in Section 2.
 The star cluster sample is described in Section 3 while the cleaning of the colour-magnitude diagrams
 (CMDs) and the derivation of the cluster parameters (size, age, metallicity) are presented in Sections 
4 and 5, respectively. Finally, we discuss the results in Section 6 and draw our main conclusions of this analysis in Section 7.

\section{Data collection and reduction}

The VMC survey strategy involves repeated observations of tiles across the Magellanic system, where 
one tile covers uniformly an area of $\sim 1.5$ deg$^2$, as a result of the mosaic of six paw-print
 images, in a given waveband with 3 epochs at $Y$ and $J$, and 12 epochs at $K_s$. Individual epochs 
have exposure times of $800$ s ($Y$ and $J$) and $750$ s ($K_s$).
The average quality of the VMC  data analysed here corresponds to  $0.34^{\prime\prime}$ pixel size,
 $0.90^{\prime\prime}$ FWHM, and 0.06 ellipticity.


To date eleven tiles in the LMC are completely observed, three of them are located in the innermost 
region of the LMC, nominally the tiles LMC 6$\_$6, 6$\_$4 and 5$\_$5.
The tile LMC 6$\_$6 (30 Doradus) is a high rate star formation region affected by large differential
 extinction (see for example Rubele et al 2012, Tatton et al. 2013), the tile LMC 6$\_$4 is located in 
the centre of the LMC Bar with high levels of crowding that could affect the capability of our tools 
to detect stars in clusters and decontaminate them by the  LMC background stars.
Star clusters in these central tiles will be analysed separately.

The tile LMC 5$\_$5 is located towards the LMC outer Bar/Bar 
region (centred at $RA=05:24:30$, $DEC= -70:48:34$ (J2000)), contains 77 catalogued clusters
with a noticeable field star crowding level and moderate extinction. 
It was completed early in the course of the survey, and we obtained 
PSF photometry of the clusters in this tile.
Consequently, we can probe our performance in reaching the Main Sequence Turnoffs (MSTOs) of
intermediate-age and relatively old clusters in crowded fields. 
Our previous experience \citep{cetal11,retal12} shows that the widest colour range of the 
$Y-K_s$ colour is best for cluster studies because it makes clearer distinguishing
different cluster Main Sequences (MSs) -particularly their turnoff regime- and the Red Giant phases, 
as well as having a higher sensitivity to reddening and metallicity
than the $Y-J$, and $J-K_s$ colours. Therefore, we mainly rely the present
analysis on the $K_s$ versus $Y-K_s$ CMDs; the $J$ versus $Y-J$ and $K_s$ versus
 $J-K_s$ CMDs being useful in order to confirm our results. 
 
The tile LMC 8$\_$8 was one of the first two fully completed VMC survey tiles, and it overlaps with 
the SEP field. The tile is centered at $RA=05:59:23$,
$Dec= -66:20:28$ (J2000) and includes 23 catalogued clusters, out of which
 two are binary clusters (KMHK\,1552 + BSDL\,3190, and KMHK\,1519 + BSDL\,3174; 
Dieball et al. 2002). The clusters catalogued by B08 located
within the limits of tiles LMC 5$\_$5 and 8$\_$8 are listed in Table 1 (see also Fig. 1). 

The tile LMC 5$\_$5 and 8$\_$8 data refers to observations acquired from November 2009 to December 2012 under
 homogeneous sky conditions since it was obtained in service mode when the sky quality met the 
requested VMC criteria (see \citet{cetal11}). 
The data were reduced with the VISTA Data Flow System pipeline, version
1.1 (VDFS; Irwin et al. 2004), and calibrated into the VISTA photometric 
system, which is close to the Vegamag system; we extracted it from the 
VISTA Science Archive (VSA; Cross et al. 2012).

For this work we perform our point-spread function (PSF) photometry on a homogenized VMC deep tile image,
that was created starting from the paw-print VMC images.
The PSF homogenized methodology consists in a convolution with a kernel of the original paw-print images
to turn different PSF shapes into a more constant and uniform PSF model on the paw-print images.
The purpose of degrading the PSF on paw-print images, to a unique PSF model, is to produce deep tile images
 with a uniform and homogeneous PSF.
As a matter of fact the paw-print images are stacked single exposures reaching a continuous
observation time of hundreds seconds,
 therefore variations of the seeing occurring over these time scales could affect the PSF shape on
the final deep tile, as a function of the position.

We performed PSF photometry on the homogenized deep tile image -created as described in Rubele et al. 
(2012)- of VMC tiles LMC 5$\_$5 and 8$\_$8, using the IRAF DAOPHOT packages \citep{s87}. The PSF model was 
created 
using $\sim$2500 stars uniformly distributed and with magnitude close to the saturation limit + 1.5 mag.
 (for the VMC survey the single paw-print saturation limits are 12.9 mag, 12.7 mag , 11.4 mag in $Y$, $J$, 
and $K_s$, respectively).         
Subsequently we used the ALLSTAR routine to perform the final PSF photometry on all the three filters 
images, and correlated the resulting catalogs using a one arcsec radius.
We checked and corrected our PSF photometry to the aperture effect using catalogs retrieved
from the VSA (Lewis et al. 2010; Cross et al. 2012)\footnote{http://horus.roe.ac.uk/vsa/}, 
for the bulk of the observed stars. We ran a large number of artificial star tests (ASTs) to 
estimate the incompleteness and error 
distribution of our data for each tile and throughout the CMD. For each region we ran  $\sim$  $20 \times 10^6$
ASTs as described in \cite{retal12}, using a spatial grid with 25 pixels width and with a 
magnitude 
distribution proportional to the square of the magnitude. This latter choice allows us to better map 
completeness and errors in the less complete regions of the CMD. Fig. 2 depicts CMDs for both tiles with
error bars coloured according to the colour scale of the completeness level.
Photometric errors of 0.10 mag were derived for stars with
 $Y$ = 19.95 mag, $J$ = 19.78 mag, and $K_s$ = 19.27 mag in the tile LMC 5$\_$5, and 
$Y$ = 21.52 mag, $J$ = 21.23 mag, and $K_s$ = 20.43 mag in the tile LMC 8$\_$8. As for the
photometry completeness we found that our data set is 50$\%$ complete at 
$Y$ = 20.6 mag, $J$ = 20.3 mag, and $K_s$ = 19.9 mag in the tile LMC 5$\_$5, and
at $Y$ = 22.1 mag, $J$ = 21.7 mag, and $K_s$ = 20.6 mag in the tile LMC 8$\_$8.

\section{The cluster sample}

We analysed a total of 98 (75 in LMC 5$\_$5 and 23 in LMC $8\_$8) candidate clusters spread over the area 
covered by the tiles considered. They are all the objects catalogued by B08 which
overlap the tile areas, in addition to BSDL\,2147 and BSDL\,2221 in LMC 5$\_$5, which
we discarded because they fall in a small tile region affected by dead pixels
 and on the edge of the tile, respectively. The studied objects range
from intermediate-age cluster candidates (age 
$\le$ 6 Gyr) to very young clusters (age $\sim$ 20 Myrs). We also confirmed that some of the
previously catalogued clusters are not indeed real stellar aggregates, but possible  cluster-like 
asterisms  (see section 6). Furthermore, 
the angular resolution of VMC made it apparent
that two catalogued clusters (KMHK747, OGLE366 in LMC 5$\_$5) are most likely compact galaxies. 
Consequently, the analysis
of the candidate clusters allowed us to attain a more robust cluster sample with genuine physical
systems in this particular field. We also refer the reader to the work by \citet{p14}
for a discussion about the completeness of the presently known star cluster population.

The confirmed clusters with ages
larger than 1 Gyr will allow us to explore overall features 
related to the star formation and chemical evolution history of the LMC. For instance, an important 
burst of cluster formation took place $\sim$ 2 Gyr ago after a 
cluster age gap \citep{p11,petal02}. On the other hand, younger clusters have been studied in the 
context of a 
variety of different astrophysical issues, like the initial mass function, the recent star 
formation rate, the early star cluster disruption \citep{dretal09,is11,dgetal13}, among others.
The accuracy of the astropysical properties derived of clusters covering a wide age range allows us 
to assess the ability of
the VMC survey in dealing with such a variety of objects, particularly those of relatively
small angular size and projected towards crowded regions such as in the LMC outer Bar.

In order to avoid mismatching between the  observed objects and the actual list of 
catalogued clusters, we first overplotted the positions of catalogued clusters 
(B08) to the deepest $K_s$ image. Thus, by using the names and the coordinates provided by 
B08, we recognised the observed clusters one by one in the $K_s$ image. Then, we searched 
such clusters in the 
Digitized Sky Survey (DSS)\footnote{The Digitized Sky Surveys were produced at the 
Space Telescope Science Institute under U.S. Government grant NAG W-2166. The images of 
these surveys are based on photographic data obtained using the Oschin Schmidt Telescope on 
Palomar Mountain and the UK Schmidt Telescope. The plates were processed into the present 
compressed digital form with the permission of these institutions.} - since they were 
originally identified from optical data - and downloaded 
15$\arcmin$$\times$15$\arcmin$ $B$ images centred on the coordinates matched by the DSS. We 
also used the SIMBAD Astronomical Database as an additional source for checking cluster 
coordinates. Finally, we compared the DSS extracted regions with the respective
ones in the $K_s$ VMC survey. We are confident of the matching procedure, particularly
when dealing with multiple cluster systems. Note that most of the observed objects
are of small angular size, typically $\sim$ 0.8$\arcmin$ in diameter ($\sim$ 11.6 pc), 
and are projected
towards relatively crowded fields with significant stellar density fluctuations. Such 
adverse physical conditions made harder not only to distinguish a star cluster from a
chance grouping of stars, but also to realiably determine the cluster centres. 

\section{Cleaning the cluster Colour-Magnitude Diagrams}

The catalogued cluster candidates appear in the sky as small concentrations of stars that
do not necessarily lead to the conclusion that such concentrations constitute real physical
systems. They may imply that we are dealing with the presence of a genuine star cluster;
a chance grouping of stars along the line-of-sight or a non uniform  distribution of 
interstellar material in that surveyed region. The CMDs of the
stars located within a region around the catalogued cluster centres are a helpful tool in 
order to assess the real entity of the objects. Nevertheless, given the significant 
fluctuations seen  in the stellar density in that part of the LMC as well as in the
luminosity function and in the colour distribution, the CMDs alone might 
lead to wrong interpretations \citep{nw00,vdm01}. In general, the extracted CMDs are the result of the 
composite stellar population distributed along the line-of-sight.

For this reason, we employed a cleaning procedure that compares the extracted CMD centred
on the cluster coordinates to four distinct CMDs built from field stars located
reasonably beyond the object but not as far as to lose the local star field signature in
terms of stellar density, luminosity function and colour distribution. As a rule, the
cluster region encompassed a circular area with a radius 3 times that of the cluster; 
the latter was taken either from a visual inspection on the deepest $K_s$ image
(where the profile disappears into the background noise) or from B08
 or from both sources combined. The four field regions were designed 
to have an equal cluster area, and were placed to the North, to the East, to the South, and to 
the West besides the cluster area. We statistically reproduced the four field CMDs
by means of box-shaped cells of different dimensions, and then used them to clean the
cluster CMD by subtracting one
star per box-shaped cell -that located within the cell and closest to its centre-, whose position
and size were defined from the field CMDs. We refer the reader to
\citet[see their Fig. 12 which illustrates the cell definition]{pb12} for a detailed
description of the field star cleaning procedure.

The method relies on the fact that some parts of the field CMD are more populated than 
others, so that by counting the number of stars within boxes of a fixed size
becomes a less effective task. In general, bigger boxes are required to
satisfactorily reproduce CMD regions with a small number of field stars, while smaller 
boxes are necessary in populous CMD regions. For instance, relatively bright field red 
giants with small photometric errors can be subtracted only if large enough boxes are used, 
allowing that a cluster CMD without such a spurious red giant features can be built.

The method allows that the cells defined on the field CMDs vary in magnitude and colour separately.
This is done by starting with a reasonable big box (($\Delta$$K_s$,$\Delta$($Y-K_s$)) =
(1.00,0.25) mag) centred on each field star and then by
getting it down in size until reaching the star closest in magnitude and that closest in
colour, respectively,
so that it results bigger in CMD regions with a small number of stars, and vice versa
(see, Fig. 3, upper-right panel).
Then, the whole amount of designed cells are plotted
over the cluster CMD and the closest stars to the cell's centres in the cluster CMD
are eliminated, independently of possible overlapped cells.

We performed the background subtraction four times per cluster, once for each field
region. When comparing the four resulting decontaminated cluster CMDs, we find stars
that have kept unsubtracted different times. The different number of times that a star 
keeps unsubtracted can then be converted in a measure of the probability of being a
fiducial feature of the cleaned region. Thus, we are able to distinguish field 
populations projected on the cluster area, i.e., those with a probability of
being a fiducial feature P $\le$ 25\%;  stars that could 
indistinguishably belong to the field or to the studied object (P = 50\%);
and stars that are predominatly found in the cleaned area (P $\ge$ 75\%)
rather than in the star field population. 

To illustrate the performance of the cleaning procedure, we show in Fig. 3 a schematic 
chart for the SL\,435 field (bottom-right panel); the extracted cluster CMD for the stars 
distributed within the circle drawn in the schematic chart (upper-left panel); a single 
field CMD for an annulus centred on the cluster, with an internal radius 3 times that of 
the circle drawn in the schematic chart and an area equal to that used to built the 
cluster CMD (upper-right panel); and the cleaned cluster CMD (bottom-left). 
We overplotted on the field CMD the cells designed for each star, which are thought
to be superimposed to the cluster CMD in order to eliminate from it one star per cell,
specifically those closest to the respective cells' centres. The colour scale 
used for the symbols in the bottom panels represents stars that statistically belong to the 
field (P $\le$ 25\%, pink), stars that might belong either to the field or to the 
cluster (P  $=$ 50\%, light blue), and stars that predominantly populate the cluster region 
(P $\ge$ 75\%, dark blue). Notice that the cluster region has more Red Clump, lower and upper
Red Giant Branch stars as well as a less populated MS than the field. A full sample of 
figures for the remaining studied objects are  provided with the on-line version of the Journal. 

It is apparent from some CMDs that the stars with the highest cluster membership 
probability (dark blue filled circles) do not define traceable cluster sequences in 
the CMD and/or are not concentrated within the circular cluster areas.
Particularly, residuals at the Red Clump (RC), Subgiant Branch, and the lower Main 
Sequence are visible. For this reason, when sizing up the clusters' reality and
estimating their fundamental parameters, we used at a time the information coming from the 
CMD and from the spatial distribution as well, i.e., we tried to make compatible the
conclusions separately drawn from the analysis of both bottom panels in the produced figures.
We also examined the cluster nature of the catalogued objects and their dimensions 
on the $K_s$ and DSS images. We usually marked on them the stars with different probabilities 
P(\%), recognized unresolved objects, and estimated the reachable limiting magnitude.

Measuring the cluster sizes proved particularly challenging. We could not estimate them by 
fitting the radial profile with King's or other profiles because most of the clusters are remarkably 
small and/or have very few possible members that make it not feasible to build stellar 
radial profiles (see next section). Instead, we adopted radii which represent a compromise 
between maximizing
the number of stars with P$>$ 75$\%$ and minimising those with P $\le$ 50$\%$ in the CMD 
and in the sky, simultaneously. The reached limiting magnitude to resolve stars also played an 
important role. Fortunately, there is a handful of clusters in our sample with available 
CMDs in the literature which served us for comparison purposes (HS\,329, 264, NGC\,1987, SL\,510). 
The central coordinates and radii of all the cluster candidates are listed in Table 1.

\subsection{Stellar profile fitting}

Our first step to determine the structural parameters of the clusters
has been to build their stellar density profiles after the field decontamination, since
 the background is not uniform
at the spatial scale of the cluster, and therefore it does not add a constant
to the surface density profile.

First, we re-determined the centre of each cluster by averaging
the coordinates of all the VMC survey stars within the ellipse given
by B08. Once the tidal radius was calculated (see below),
the centre was re-determined. The procedure was repeated a few
times, until the position was stable to within 1/1000\,deg.
Our final coordinates agree well with B08's values to within
(5$\pm$3)\,arcsec.


Next, we fitted the ``classical'' \citet{k62} and EFF
\citep{eetal87} profiles, given by:

\begin{equation}
n(r) = k \times \{1/[1+(r/r_0)^2)]^{1/2}-(1/[1+(r_t/r_0)^2]^{1/2}\}^2 + \phi_K
\end{equation}

and
\begin{equation}
n(r) = n_0 \times \{1+(r/a)^2\}^{-\gamma/2} + \phi_E,
\end{equation}
respectively. Here:
$n(r)$ is the number of stars per unit area as a function of
the radius $r$;
$k$ and $n_0$ are central projected stellar number densities;
$r_0$ is the King's radius;
$a$ is a core parameter, related to the core radius $r_c$:
\begin{equation}
r_c = a \times (2^{2/\gamma}-1)^{1/2};
\end{equation}
$r_t$ is the tidal radius;
$\phi_K$ and $\phi_E$ are background stellar number densities.
We also calculated the concentration parameter $c$:
\begin{equation}
c = {\rm log}(r_t/r_0).
\end{equation}

Table 2 presents the results from these fits for 30 clusters for
which the fitting method converged,
while Fig. 4 shows an example of radial and fitted profiles. 
Most of the clusters in the tile LMC 5\_5 are of small angular size,
contain a few number of stars, and are projected towards
relatively crowded star fields as well, which mainly caused the
fitting procedure not to converge.
For comparison, we list the cluster sizes from \citet{betal08} and
\citet{ketal90}, converted into arcsec.
Finally,
we explored the ellipticity of the clusters, fitting
ellipses to the radial profile, but the typical deviation from a
circular shape was $\leq$10\%, comparable to the uncertainty in
the cluster sizes, so we refrained from further investigation of
the 2-dimensional cluster shapes.

Unfortunately, for many clusters (and particularly for the smaller ones)
we obtain values of the tidal radius $r_t$ that
are unreliable and so large ($\geq$ 100 arcsec) that become
meaningless to be exploited for the isochrone fitting; this could
be due to the fact that many clusters in the sample
are small and have not enough stars to build reliable stellar radial profiles.

\section{Cluster fundamental parameters}

In order to estimate ages for the catalogued cluster sample, we first adopted
appropriate reddening from the available reddening maps, and distance modulus equal to that of
the LMC, and  then fitted theoretical isochrones covering a wide range in age 
and metallicity to the probable (P $\ge$ 75\%) cluster members on the $K_s$ vs. $Y - K_s$ CMDs. 
The colour excesses and the distance modulus played
a double role. On the one hand, they made the number of variables to be considered in the
theoretical isochrone fits smaller. Instead of simultaneously varying four parameters, we only
looked for the age and the metallicity of the isochrone which best matched the cluster features.
On the other hand, they served as an external control of the zero point of our photometry. 
Fortunately, in all the analysed CMDs, the Zero Age Main Sequence (ZAMS) satisfactorily lies over the
observed cluster MS. 

The estimation of cluster reddening values was made by taking
advantages of the Magellanic Clouds extinction values based on red clump stars 
photometry provided by the OGLE collaboration \citep{u03} as described in \citet{hetal11}.
These $E(B-V)$ colour excesses are listed in Table 1.
They resulted in average (0.02 $\pm$ 0.01) mag smaller than those obtained 
by \citet{cetal03}. As for the \citet{sfd98} 
full-sky maps from 100-$\mu$m dust emission, we decided not to use them since the authors found that 
deviations are coherent in the sky and are especially conspicuous in regions of saturation of H\,I emission 
towards denser clouds and of formation of H$_2$ in molecular clouds \citep[see also,][]{petal03,petal08}. 
We note that the
 small angular size
of the studied clusters does not allow to trace reddening variations in any extinction map. 

We  adopted for all the clusters the value of the LMC distance modulus 
$(m-M)_o$ = 18.49 $\pm$ 0.09 (49.90$^{-2.04}_{+2.10}$ kpc) \citep{dgetal14}, and an average depth 
for the LMC disk of 
(3.44 $\pm$ 1.16) kpc \citep{ss09}. Bearing in mind that any cluster of 
the  present sample could be placed in front of, or behind the LMC, we conclude that the 
difference in apparent distance modulus could be as large as $\Delta({K_s}-M_{K_s})$ $\sim$ 0.3 mag, if a 
value of 18.49 mag is adopted for the mean LMC distance modulus. Given the fact that we estimate an uncertainty 
of the 
order of 0.3 mag when 
adjusting the isochrones to the cluster CMDs in magnitude, our simple assumption of adopting a unique value
 for 
the distance modulus of all the clusters should not dominate the error budget in our final results. 
In fact, 
when overplotting the ZAMS on the observed cluster CMDs, previously shifted by the $E(B-V)$ values
of Table 1 and by $(m-M)_o$ = 18.49, we generally found an excellent match. 

The ages and metallicities have complex and intertwined effects on the shape of the 
cluster's CMD. The 
distinction is mainly evident for the evolved RC and Red Giant Branch (RGB) phases. ZAMSs are often less 
affected by
metallicity effects and can even exhibit imperceptible variations for a specific metallicity range 
within the expected photometric errors. Since the LMC chemical evolution  has mostly taken place 
within a constrained metallicity range during the last 3 Gyr, we simply used [Fe/H] values of 
-0.4 dex and -0.7 dex \citep{pg13}. Further higher metallicity resolution would lead to 
negligible changes in the isochrones overplotted on the cluster CMDs due to the dispersion of the stars.
We took advantage of the available theoretical isochrones
computed for the VISTA photometric system to estimate  cluster ages. We used  recent isochrones 
calculated  by the Padova group \citep{betal12}.
We then selected a set of isochrones, along with the equations $E(Y-K_s)$ = 0.84$\times$$E(B-V)$ and 
$K_s$ = $M_{K_s}$ + $(m-M)_o$ + 0.372$\times$$E(B-V)$ with $R_V$ = 3.1 \citep{cetal89,getal13}, 
and superimposed them on the cluster CMDs, once they were 
properly shifted by the corresponding $E(B-V)$ colour excesses and by the LMC apparent distance 
modulus. In the matching procedure, we used a subset of isochrones for each metallicity 
level, ranging from $\Delta$(log($t$/yr)) = -0.3 to +0.3 around a first rough age estimate. Finally, 
we adopted as the cluster age/metallicity the ones corresponding to the isochrone which best reproduced the cluster 
main features in the CMD. The presence of RCs and/or RGBs in some cluster CMDs made the fitting procedure 
easier. Table 1 lists the resulting age and metallicity values, while the bottom left panel in Fig. 3  (and for 
the complete sample of clusters in Appendix A) show the corresponding isochrones superimposed to the 
cluster CMDs.
We found that
isochrones bracketing the derived mean age by $\Delta$(log($t$/yr)) = $\pm$0.1 reasonably represent the overall age 
uncertainty due to the observed dispersion in the cluster CMDs, as shown in Fig. 3 (bottom-left panel). 
Although in some cases the age dispersion is smaller than $\Delta$(log($t$/yr)) = 0.1, we prefer to keep
the former value as an upper limit of our error budget \cite[among others]{p10,petal11,p14}. 
On the other hand, by assuming that both used metallicity values satisfy 
$\sigma$([Fe/H]$_{\rm 1}$=-0.4 dex) + $\sigma$([Fe/H]$_{\rm 2}$=-0.7 dex) $\ge$ 
$|$ [Fe/H]$_{\rm 1}$ - [Fe/H]$_{\rm 2}$ $|$, and $\sigma$([Fe/H]$_{\rm 1}$ = $\sigma$([Fe/H]$_{\rm 2}$, 
we adopted metallicity error of $\sigma$([Fe/H]) = 0.15 dex.
In the case of SL\,441, we found that isochrones with [Fe/H] = -0.4 dex do not
satisfactorly match the cluster CMD as compared to those with solar metal content. The rather
high metallicity for the LMC makes SL\,441 interesting for further studies; particularly because there is 
no previous detailed study on this object. Nevertheless, 
since we are able to distinghish between isochrones with [Fe/H] = -0.3 dex and 0.0 dex, we also
assume for this cluster a metallicity error of 0.15 dex.

\section{Discussion}

We finally estimated the ages of 65 clusters out of the 98 studied objects; 19 of them have some previous 
age/metallicity estimates. We have included this latter information in the last
column of Table 1 and plotted the age differences in Fig. 5. \citet{getal10}
have used data from the Magellanic Cloud Photometric Surveys \citep{zetal02} to build 
cluster CMDs and to derive their ages from theoretical isochrone fits. Although they mention that field 
contamination is a severe effect in the extracted cluster CMDs and therefore influences the age 
estimates, no
decontamination from field CMDs were carried out. Consequently, their large age errors could reflect  
the composite LMC stellar populations. Thus, as an example, they estimated for HS\,232 an age of log($t$/yr)=
9.2$\pm$0.1, whereas from our analysis we could not confirm the object like a possible star cluster.
Likewise, from a total of 14 clusters in common, we found a difference of
$|$log($t$/yr)$_{\rm glatt}$ - log($t$/yr)$_{\rm our}$$|$ = 0.3 $\pm$ 0.4 (absolute values). For the remaining 5
clusters with previous age estimates, we found an excellent agreement (see Table 1). 

As for the metallicity estimates,  \citet{getal10} adopted a value of [Fe/H] = -0.4 dex for the
14 clusters in common, which is in excellent agreement with our values except for SL\,435, 
for which we used a more metal-poor isochrone ([Fe/H]=-0.7 dex). We think that the difference
in metallicty for SL\,435 could be due to field contamination effects as mentioned above, since the
cluster age also differs significantly. Others three clusters with metallicity values published in the 
literature are NGC\,1987, NGC\,2010, and SL\,510. \citet{metal09} obtained the best
isochrone fit to the NGC\,1987 CMD using Z= 0.010 ([Fe/H]=-0.3 dex), while  \citet{goetal10} and \citet{p12}
assumed a metallicity of [Fe/H] = -0.4 dex for NGC\,2010 and SL\,510, respectively. As can be seen,
our present values are in excellent agreement with those previously published. On the other hand,
the percentage of clusters with [Fe/H] = -0.7 dex resulted in nearly the same value
for both tiles ($\approx$ 20\%), as well as the mean cluster metallicities ([Fe/H] = -0.45 $\pm$ 0.15 dex);
the number of more metal-rich clusters being higher in tile LMC 5$\_$5 than in tile LMC 8$\_$8.

SL\,528 and OGLE\,545 are two objects located very close to each other in the sky (angular separation 
$\approx$ 0.55$\arcmin$) whose decontaminated CMDs look pretty similar. They do not show the cluster MSs, but
exhibit visible RCs. We interpret this effect either as coming from our not deep enough photometry
of intermediate-age clusters (log($t$/yr)$>$9) or as dealing with cluster-like asterisms,
i.e, like a statistical fluctuation of the field or a low-absorption window \citep{bb11}.
BSDL\,1182 also has a decontaminated CMD similar to those of SL\,528 and OGLE\,545.
However, in this case we think that we are dealing with a group of RC stars or with an
unresolved star cluster.  We recall that in order to asses the objects' reality
we used at a time the information coming from the 
CMDs and from the spatial distributions as well.

As far as we are aware, the most recent catalogue of LMC star clusters
which puts all the previous ones together is that of B08. Although 
it is expected that most of the catalogued objects are indeed genuine physical systems, it 
was beyond the scope of B08 to verify the physical nature of such faint objects.
The task of cleaning cluster catalogues from non-physical systems or asterisms
is far from being an exciting job.
For this reason studies concluding about the 
asterism 
or overdensity nature of faint objects in the Clouds are rare or absent \citep{pb12}.
However such analysis would be very important and would be
required for any statistical analysis of the cluster 
formation and disruption rates, the cluster spatial, age and metallicity distributions, etc. is attempted. 
Since B08's catalogue was compiled from  previously existing catalogues built on the 
basis of star counts, either by visually inspecting photographic plates 
\cite[for example]{b75,h86,bs95} or by
automatic algorithmic searches \cite[for example]{petal99}, we should not rule out the possible 
occurrence of such asterisms.

Indeed, we classified 19 of the studied objects as possible non-clusters. For them, although 
apparent concentrations of stars in a typically 1$\arcmin$ a side angular region can be observed in the
$K_s$ images, a careful inspection of the resulting spatial distributions and the decontaminated CMDs for 
stars with P($\%$) $>$ 75 did not allow us to firmly conclude that they are genuine physical systems.
However, the present analysis tools applied to faint poorly populated clusters or candidates in the
LMC points to the need of  higher spatial resolution and deeper observations with e.g. the 8m class 
telescopes. For 8 objects, we could not find a clear star concentration in the $K_s$ images either, so 
that we quoted them as cluster-like asterisms in Table 1. Finally, we found two previously catalogued
single star clusters to be unresolved background galaxies  on the basis of isophote analyses and the 
comparison of radial profiles of a sample of objects in the images (KMHK\,747, OGLE\,366) and a triple 
cluster system 
(BSDL\,2144: (RA, Dec) $\approx$ (05:31:10.0,-71:08:00.0), (05:31:08.0,-71:07:50.0), and 
(05:31:05.0,-71:07:55.0)), respectively. Nevertheless, better spatially resolved images are needed
in order to give more conclusive results. Instead of building their schematic charts and CMDs, we provide here
with the respective $K_s$ images in order to have a better judgement of them (see Fig. 6). We have superimposed
isophotes curves which highlight the unresolved nature of these objects. For comparison
purposes we include an enlargment of the $K_s$ image centred on the cluster SL\,435 in the bottom-right panel
of Fig. 6 (see also Fig. 3).

\citet{p14} showed that there exist some variations in the LMC star cluster frequency (CF)
 in terms of cluster spatial distribution. Particularly he found that  30 Doradus turns out
 to be the region with the
highest relative frequency of the youngest clusters, while the log($t$/yr) = 9-9.5 (1-3 Gyr) age
range is characterized by cluster formation at a higher rate in the inner regions than in the
outer ones. In Fig. 7 we compare the CF we obtained for the tile LMC 5$\_$5 to those obtained by
\citet{p14} for the
Bar and the outer Bar, since this tile covers regions of both structures \citep{hz09,metal14}.
When building the CF we took into account the same precepts outlined in \citet{p14}, i.e., the
influence of adopting arbitrary age bins, as well as
the fact that each age value is associated to an uncertainty which allows the age value to fall 
centred on an age bin or outside it. In practice, we varied the bin size based on the average 
error of the age of the clusters that fall in each bin, thus tracing the variation of the age 
uncertainties along the whole age range. In addition, even though the age bins are set to match 
the age errors, any individual point in the CF may fall into the respective age bin
or either of the two adjacent bins. This happens when an age point
does not fall in the bin centre and, owing to its errors, has the chance
of falling outside it.  For this reason, we weighed the contribution of each age
value - a segment with size 2$\sigma$(age) - to each one of the age bins occupied by it, so that the sum
of all the weights equals unity.
We  performed thus a robust procedure which achieves to take into account both effects: the 
age bin size and the age errors. Since the total number of clusters in the tile
and in the sample used by Piatti \cite{p14} is different, we
normalised the CF to the total number of clusters employed, for comparison purposes. Note that
we are interested in comparing the slope or changes in the CF rather than the total number of clusters
formed per age interval.
Although tile LMC 5$\_$5 encompasses a small portion of the LMC outer Bar/Bar (see Fig. 1), we 
found that its CF resulted to be comparable to those of the outer Bar and the Bar.  This
results might reflect that the different parts of the LMC outer Bar/Bar have behaved in a similar
way forming star clusters during the last ~ 1-2 Gyr. Note, however, that
we did not measure ages of clusters older than $\sim$  log(t/yr) = 9.6 (4 Gyr) which, in turn, 
might suggest that old clusters are not as homogeneously distributed as those younger ones in
terms of CF in the inner LMC regions.

\section{Conclusions}

In this work we analyse CMDs of catalogued star clusters located in the 
LMC from a $YJK_s$ photometric dataset obtained by the VISTA VMC collaboration. We focused on tiles
LMC 5$\_$5 and 8$\_$8 because they are among of the firstly completed by the VMC survey for 
which we obtained PSF photometry. Since they are respectively located towards a LMC outer Bar/Bar and
SEP regions, we could assess the performance of estimating ages for the oldest clusters
observed (i.e., limiting magnitude reached) in relatively crowded and uncrowded fields.
We analysed a total of 98 catalogued clusters of small angular size, typically $\sim$ 11.6 pc in 
diameter

We applied a subtraction procedure developed by \citet{pb12} to statistically clean the cluster CMDs
 from field star contamination 
in order to disentangle cluster features from those belonging to their surrounding fields.
The employed technique makes use of variable cells in order to reproduce the field CMD as closely as
possible. 

From matching theoretical isochrones computed for the VISTA system to the cleaned cluster 
CMDs we estimated ages and metallicities. When adjusting a subset of isochrones we took into 
account the LMC distance modulus and the individual star cluster colour excesses. We finally estimated the
 ages 
of 65 clusters out of the 98 studied objects, which resulted to be in the age range 7.3 $<$ log($t$/yr)
 $<$ 9.55.
This cluster sample will be part of the cluster data base that the VMC survey will produce in order to
homogeneously study the overall cluster formation history throughout the Magellanic system.

We also classified 19 of the studied objects as possible non-clusters (e.g., chance grouping of stars)
since a careful inspection of the
 resulting 
spatial distributions and the decontaminated CMDs for stars with probabilities of being a fiducial cluster 
feature higher
than 75$\%$ did not allow us to firmly conclude that they are genuine physical systems. Other
two objects were classified as possible cluster-like asterisms and another one as an unresolved cluster.
For other 8 objects, we could not find a clear star concentration in the $K_s$ images either, so 
that we quoted them as cluster-like asterisms. Finally, we found two previously catalogued
single star clusters to be unresolved background galaxies (KMHK\,747, OGLE\,366) and a triple cluster system 
(BSDL\,2144), respectively.

\begin{acknowledgements}
We thank the Cambridge Astronomy Survey Unit (CASU) and the Wide Field Astronomy Unit (WFAU) in Edinburgh 
for providing calibrated data products under the support of the Science and Technology Facility 
Council (STFC) in the UK.
This research has made use of the SIMBAD database,
operated at CDS, Strasbourg, France. 
This work was partially supported by the Argentinian institutions CONICET and
Agencia Nacional de Promoci\'on Cient\'{\i}fica y Tecnol\'ogica (ANPCyT). 
R.G. is a Postdoctoral Fellow - Pegasus of the Fonds Wetenschappelijk Onderzoek (FWO) - Flanders.
RdG acknowledges research support from the National Natural Science
Foundation of China (NSFC) through grant 11373010. GC acknowledges research support from 
PRIN-INAF 2010 (P.I.G. Clementini). B.-Q.F. is the recipient of a John Stocker Postdoctoral
 Fellowship from the Science and Industry Endowment Fund.  We thank the anonymous referee whose 
comments and suggestions allowed us to improve the manuscript.

\end{acknowledgements}

\clearpage

\setcounter{table}{0}
\begin{table*}
\caption{Fundamental parameters of the studied LMC cluster candidates.}
\centering
\begin{tabular}{lccccccc}\hline\hline
ID & R.A.       & Dec.       & r            & E(B-V) &  log($t$/yr)  & [Fe/H] & Note \\ 
   & ($\degr$)  & ($\degr$)  & ($\arcmin$)  & (mag)  &            & (dex)  & \\\hline
\multicolumn{8}{c}{LMC 5$\_$5} \\\hline
BSDL\,1504    & 80.775 & -71.422 & 0.4  & 0.065 & 9.55 &-0.7 &  \\
BSDL\,1355    & 80.342 & -70.900 & 0.4  & 0.064 & 9.50 &-0.7 &     \\
BSDL\,1341    & 80.249 & -70.318 & 0.5  & 0.085 & 9.45 &-0.7 &        \\
SL\,435       & 80.875 & -71.428 & 0.4  & 0.065 & 9.40 &-0.7 &   8.70$\pm$0.20, -0.4 (1) \\
KMHK\,897     & 81.492 & -70.459 & 0.7  & 0.085 & 9.40 &-0.7 &      \\
KMHK\,835     & 80.533 & -71.175 & 0.5  & 0.064 & 9.40 &-0.7 &      \\
HS\,329       & 82.442 & -71.001 & 0.4  & 0.070 & 9.30 &-0.7 &   9.25$\pm$0.04 (4) \\
KMHK\,750     & 79.577 & -70.721 & 0.4  & 0.086 & 9.30 &-0.4 & \\
SL\,472       & 81.562 & -70.380 & 0.4  & 0.085 & 9.20 &-0.7 &   \\
BSDL\,1672    & 81.375 & -71.087 & 0.25 & 0.070 & 9.20 &-0.4 &      \\
HS\,323       & 82.215 & -70.207 & 0.6  & 0.071 & 9.20 &-0.4 & \\
HS\,264       & 80.805 & -70.778 & 0.4  & 0.082 & 9.10 &-0.4 &  9.20$\pm$0.04 (4) \\
KMHK\,997     & 82.576 & -70.420 & 0.5  & 0.076 & 9.10 &-0.4 & \\
NGC\,1987     & 81.821 & -70.738 & 0.8  & 0.080 & 9.00 &-0.4 &    9.05$\pm$0.05, 0.010 (3)  \\
SL\,389       & 79.905 & -71.211 & 0.7  & 0.064 & 8.90 &-0.4 & \\
HS\,214       & 79.460 & -70.801 & 0.3  & 0.086 & 8.90 &-0.4 &  8.15$\pm$0.04, -0.4 (1) \\ 
KMHK\,907     & 81.553 & -70.981 & 0.3  & 0.070 & 8.80 &-0.4 & \\
HS\,238       & 80.033 & -70.156 & 0.3  & 0.100 & 8.80 &-0.4 & \\
SL\,441       & 80.925 & -71.037 & 0.5  & 0.064 & 8.80 & 0.0 & \\
BSDL\,2123    & 82.775 & -70.168 & 0.5  & 0.071 & 8.80 &-0.4 & \\
SL\,399       & 80.087 & -70.768 & 0.5  & 0.064 & 8.50 &-0.4 &  8.30$\pm$0.05, -0.4 (1) \\
KMK\,88$\_$55   & 80.923 & -70.096 & 0.3  & 0.084 & 8.50 &-0.4 &  \\
HS\,304       & 81.814 & -71.172 & 0.4  & 0.074 & 8.50 &-0.4 &  \\
SL\,406       & 80.258 & -70.873 & 0.3  & 0.064 & 8.45 &-0.4 &  8.50$\pm$0.05, -0.4 (1) \\
SL\,395       & 79.996 & -70.665 & 0.3  & 0.083 & 8.40 &-0.4 &  8.30$\pm$0.20, -0.4 (1) \\
HS\,324       & 82.229 & -71.120 & 0.3  & 0.070 & 8.40 &-0.4 &  \\
SL\,364       & 79.423 & -71.066 & 0.3  & 0.094 & 8.30 &-0.4 &  8.00$\pm$0.20, -0.4 (1) \\
NGC\,1943     & 80.623 & -70.155 & 0.5  & 0.085 & 8.30 &-0.4 &  8.35$\pm$0.05, -0.4 (1) \\
SL\,487e      & 81.788 & -71.023 & 0.4  & 0.070 & 8.30 &-0.4 &  \\
KMHK\,764     & 79.720 & -70.602 & 0.4  & 0.091 & 8.30 &-0.4 & \\
SL\,431       & 80.800 & -70.2805 & 0.4 & 0.050 & 8.30 &-0.4 & \\
NGC\,2010     & 82.641 & -70.819 & 0.5  & 0.078 & 8.20 &-0.4 &   8.20$\pm$0.05, -0.4 (2)\\
SL\,510       & 82.340 & -70.579 & 0.5  & 0.076 & 8.10 &-0.4 &   8.10$\pm$0.10, -0.4 (5)\\ 
KMHK\,999     & 82.553 & -71.557 & 0.3  & 0.076 & 8.00 &-0.4 &  \\
BSDL\,1949    & 82.345 & -70.237 & 0.4  & 0.071 & 7.70 &-0.4 &  \\
BSDL\,1876    & 81.977 & -71.522 & 0.2  & 0.076 & 7.50 &-0.4 &  8.20$\pm$0.40, -0.4 (1) \\
BSDL\,2008    & 82.460 & -71.068 & 0.3  & 0.070 & 7.50 &-0.4 &  \\
SL\,539       & 82.733 & -70.695 & 0.4  & 0.078 & 7.50 &-0.4 &  7.40$\pm$0.20, -0.4 (1) \\
BSDL\,2199    & 82.942 & -70.253 & 0.4  & 0.071 & 7.50 &-0.4 &  \\
KMHK\,979     & 82.412 & -70.986 & 0.4  & 0.070 & 7.30 &-0.4 &  7.30$\pm$0.40, -0.4 (1) \\
\hline
\end{tabular}
\end{table*}

\setcounter{table}{0}
\begin{table*}
\tiny
\begin{tabular}{lccccccc}\hline\hline
ID & R.A.       & Dec.       & r            & E(B-V) &  log($t$/yr)  & [Fe/H] & Note \\ 
   & ($\degr$)  & ($\degr$)  & ($\arcmin$)  & (mag)  &            & (dex)  & \\\hline
BSDL\,1955    & 82.332 & -71.031 & 0.3  & 0.070 & 7.30 &-0.4 &  7.30$\pm$0.40, -0.4 (1) \\
BSDL\,1980    & 82.387 & -70.994 & 0.5  & 0.070 & 7.30 &-0.4 &  7.30$\pm$0.40, -0.4 (1) \\
BSDL\,1830    & 81.892 & -70.614 & 0.2  & 0.076 & 7.30 &-0.4 &  7.50$\pm$0.60, -0.4 (1) \\
BSDL\,1875    & 81.976 & -71.547 & 0.3  & 0.076 & 7.30 &-0.4 &  8.60$\pm$0.60, -0.4 (1) \\
SL\,528       & 82.670 & -70.223 & 0.2  & 0.071 & $>$9.00 & ...  & cluster-like asterism?\\
OGLE\,545     & 82.663 & -70.217 & 0.2  & 0.071 & $>$9.00 & ...  & cluster-like asterism?\\
BSDL\,1182    & 79.568 & -71.435 & 0.15 & 0.065 &  ...       &  ...    & unresolved cluster? \\
HS\,232       & 80.155 & -70.964 & 0.2  & 0.064 & ...        &  ...     &possible non-cluster;  9.20$\pm$0.10, -0.4 (1) \\
HS\,265       & 80.817 & -70.234 & 0.25 & 0.085 & ...        &  ...     &possible non-cluster  \\
BSDL\,1790    & 81.720 & -70.211 & 0.4  & 0.085 &  ...       &   ...    &possible non-cluster  \\
OGLE\,534     & 82.515 & -70.126 & 0.2  & 0.063 & ...        &   ...    &possible non-cluster  \\
OGLE\,536     & 82.517 & -70.205 & 0.2  & 0.071 &  ...       &   ...    &possible non-cluster \\
KMHK\,801     & 80.114 & -70.450 & 0.3  & 0.085 &  ...       &   ...    &possible non-cluster  \\
HS\,295       & 81.524 & -70.092 & 0.3  & 0.085 &   ...      &   ...    &possible non-cluster \\
HS\,282       & 81.318 & -70.099 & 0.3  & 0.085 &  ...       &   ...    &possible non-cluster  \\
BSDL\,2196    & 82.932 & -70.201 & 0.2  & 0.071 &  ...       &   ...    &possible non-cluster  \\
BSDL\,2063    & 82.618 & -70.560 & 0.5  & 0.076 &  ...       &   ...    &possible non-cluster   \\
OGLE\,471     & 81.573 & -70.222 & 0.3  & 0.085 &  ...       &  ...     &possible non-cluster  \\
HS\,345       & 82.956 & -70.287 & 0.25 & 0.071 &  ...       &   ...    &possible non-cluster  \\
KMHK\,819     & 80.333 & -71.407 & 0.3  & 0.065 & ...        &  ...     &possible non-cluster  \\
HS\,342       & 82.940 & -70.308 & 0.3  & 0.071 &  ...       &  ...     &possible non-cluster   \\
SL\,542       & 82.825 & -70.216 & 0.4  & 0.071 &  ...       &   ...    &possible non-cluster  \\
BSDL\,1645    & 81.336 & -71.196 & 0.4  & 0.074 & ...        &  ...     &possible non-cluster   \\
GKK\,O102     & 82.871 & -70.757 & 0.3  & 0.078 & ...        &   ...    &possible non-cluster  \\
SL\,439       & 81.010 & -70.174 & 0.3  & 0.084 & ...        &   ...    &possible non-cluster   \\
BSDL\,2202    & 82.897 & -70.735 & 0.5  & 0.078 &  ...       &   ...    &cluster-like asterism  \\
GKK\,O15      & 80.550 & -71.319 & 0.5  & 0.065 &  ...       &   ...    &cluster-like asterism\\
KMHK\,740     & 79.449 & -71.157 & 0.4  & 0.064 & ...        &   ...    &cluster-like asterism \\
OGLE\,542     & 82.642 & -70.197 & 0.4  & 0.071 & ...        &   ...    &cluster-like asterism\\
HS\,286       & 81.408 & -70.262 & 0.2  & 0.085 &  ...       &   ...    &cluster-like asterism\\
GKK\,O119     & 80.627 & -70.360 & 0.4  & 0.085 & ...        &  ...     &cluster-like asterism\\
GKK\,O100     & 82.753 & -70.920 & 0.5  & 0.078 &  ...       &   ...    &cluster-like asterism \\
BSDL\,2144    & 82.785 & -71.131 & ...  & 0.070 &  ...       &   ...    &possible triple system \\
KMHK\,747     & 79.519 & -71.269 & ...  & 0.065 &  ...       &   ...    &possible galaxy       \\
OGLE\,366     & 80.033 & -70.144 & ...  & 0.100 &  ...       &   ...    &possible galaxy       \\\hline
\multicolumn{8}{c}{LMC 8$\_$8} \\\hline

 KMHK\,1592 &  90.375 & -66.987 & 0.8   & 0.042 &   9.8  &   -0.7&                \\
 KMHK\,1521 &  88.742 & -67.114 & 0.5   & 0.050 & 9.5   & -0.4    &                \\
 KMHK\,1585 &  90.212 & -66.913 &  0.5   & 0.042 &   9.3  &   -0.7&                \\
 KMHK\,1578 &  89.991 & -66.443 &  0.4   & 0.035 &   9.25 &   -0.7&                \\
 KMHK\,1552 &  89.468 & -65.950 &  0.2   & 0.034 &   9.2  &   -0.7&                \\
 KMHK\,1609 &  90.795 & -65.675 &  0.4   & 0.037 &   9.2  &   -0.4&                \\
 KMHK\,1577 &  89.952 & -66.770 &  0.5   & 0.042 &   9.2  &   -0.4&                \\
 KMHK\,1623 &  91.140 & -66.442 &  0.5   & 0.040 &   9.2  &   -0.4&                \\
 KMHK\,1567 &  89.703 & -66.059 &  0.5   & 0.034 &   9.1  &   -0.4&                \\
 KMHK\,1555 &  89.466 & -66.401 &  0.5   & 0.035 &   9.1  &   -0.4&                \\
 KMHK\,1597 &  90.541 & -65.776 &  0.4   & 0.034 &   9.1  &   -0.4&                \\
 KMHK\,1600 &  90.619 & -66.920 &  0.4   & 0.037 &   9.1  &   -0.4&                \\
 KMHK\,1611 &  90.835 & -66.126 &  0.3   & 0.037 &   9.1  &   -0.4&                \\
 BSDL\,3174 &  88.685 & -66.715 & 0.3   & 0.043 & 9.1   & -0.4    &                \\
 KMHK\,1568 &  89.695 & -66.844 &  0.3   & 0.042 &   9.0  &   -0.4&                \\
 KMHK\,1510 &  88.642 & -65.740 & 0.5   & 0.035 & 9.0   & -0.4    &                \\
 LW\,334    &  89.001 & -66.288 &  0.4   & 0.040 &   8.9  &   -0.4&                \\
 BSDL\,3188 &  89.325 & -66.273 &  0.5   & 0.035 &   8.9  &   -0.4&                \\
 KMHK\,1519 &  88.729 & -66.714 & 0.5   & 0.043 & 8.9   & -0.4    &                \\
 KMHK\,1516 &  88.709 & -65.838 & 0.4   & 0.035 & 8.8   & -0.4    &                \\
 KMHK\,1589 &  90.329 & -66.854 &  0.6   & 0.042 &   8.7  &   -0.4&                \\      
BSDL\,3190 &  89.472 & -65.945 &  0.15  & 0.034 &   ...  &     ...& asterism                \\
 KMHK\,1607 &  90.714 & -66.660 &  0.3   & 0.039 &   ...  &   ...  &  possible non-cluster              \\
\hline

\end{tabular}
\medskip

\noindent Note: (1) Glatt et al. 2010; (2) Gouliermis et al. 2010; (3) Milone et al. 2009;
(4) Piatti 2011; (5) Piatti 2012.

\end{table*}

\setcounter{table}{1}
\begin{table*}[ht]
\caption{Elson-Fall-Freeman and King's profile fitting results.}
\label{table2}
\scriptsize
\begin{tabular}{@{ }l@{ }r@{ }r@{ }r@{ }r@{ }r@{ }r@{ }r@{ }r@{ }r@{ }}
\hline\hline
Cluster      &\multicolumn{3}{c}{Elson-Fall-Freeman profile}&\multicolumn{3}{c}{King profile}              &\multicolumn{2}{c}{Bica}& Kon- \\ 
ID           & $\gamma$    & $a$           & $r_{\sc c}$   & r$_{\sc 0}$  & $r_{\sc t}$    & $c$           & Maj   & Min  & tizas  \\
             &             & arcsec        & arcsec        & arcsec       & arcsec         &               & \multicolumn{3}{c}{arcsec} \\
\hline
\multicolumn{10}{c}{LMC 5$\_$5} \\\hline
BSDL\,1341   &   ...        &       ...      &      ...       &     ...       &        ...      &     ...      &    33 &   30 &  ...   \\
BSDL\,1355   &   ...        &       ...      &      ...       &     ...       &        ...      &     ...      &    36 &   33 &  ...   \\
BSDL\,1504   &   ...        &       ...      &      ...       &     ...       &        ...      &     ...      &    30 &   24 &  ...   \\
BSDL\,1672   &   ...        &       ...      &      ...       &     ...       &        ...      &     ...      &    36 &   33 &  ...   \\
BSDL\,2123   &   ...        &       ...      &      ...       &     ...       &        ...      &     ...      &    66 &   51 &  ...   \\
HS\,214      & 0.4$\pm$0.1 &  0.6$\pm$ 0.2 &  2.6$\pm$ 1.8 &     ...       &        ...      &     ...      &    27 &   21 &  20.1 \\
HS\,238      &   ...        &       ...      &      ...       &     ...       &        ...      &     ...      &    48 &   39 &  ...   \\  
HS\,304      &   ...        &       ...      &      ...       &     ...       &        ...      &     ...      &    60 &   60 &  60.3 \\  
HS\,323      &   ...        &       ...      &      ...       &     ...       &        ...      &     ...      &    39 &   30 &  ...   \\
HS\,324      &   ...        &       ...      &      ...       &     ...       &        ...      &     ...      &    48 &   39 &  40.2 \\  
HS\,329      & 0.3$\pm$0.1 &  1.8$\pm$ 0.8 & 17.0$\pm$23.2 &  1.4$\pm$0.1 &   6.7$\pm$ 0.1 & 0.7$\pm$0.1 &    45 &   39 &  46.8 \\
KMHK\,750    & 0.4$\pm$0.1 &  0.9$\pm$ 0.3 &  5.6$\pm$ 3.0 &  2.8$\pm$0.1 &  13.0$\pm$ 0.1 & 0.7$\pm$0.1 &    48 &   36 &  33.6 \\
KMHK\,835    & 0.3$\pm$0.1 &  2.1$\pm$ 0.6 & 17.2$\pm$ 9.9 &  4.0$\pm$0.9 & 105.1$\pm$18.0 & 1.4$\pm$0.2 &    54 &   48 &  40.2 \\
KMHK\,897    &   ...        &       ...      &      ...       &     ...       &        ...      &             &    36 &   30 &  26.7 \\
KMHK\,907    & 0.8$\pm$0.1 &  1.0$\pm$ 0.3 &  2.2$\pm$ 0.7 &  0.8$\pm$0.1 &  13.0$\pm$ 1.7 & 1.2$\pm$0.1 &    33 &   30 &  40.2 \\
KMHK\,997    &   ...        &       ...      &      ...       &     ...       &        ...      &     ...      &    54 &   39 &  40.2 \\
KMK\,88\_55   &  ...         &      ...       &     ...        &    ...        &       ...       &    ...       &    36 &   30 &  ...   \\
NGC\,1987    & 0.3$\pm$0.1 &  3.7$\pm$ 1.0 & 32.5$\pm$19.0 &  2.6$\pm$0.1 &  17.5$\pm$ 0.3 & 0.8$\pm$0.1 &   102 &  102 &  98.4 \\
SL\,389      & 0.8$\pm$0.1 &  6.8$\pm$ 1.4 & 13.4$\pm$ 4.6 &  3.3$\pm$0.1 &  23.8$\pm$ 0.6 & 0.9$\pm$0.1 &    96 &   78 & 100.5 \\
SL\,399      & 0.6$\pm$0.1 &  1.2$\pm$ 0.2 &  3.5$\pm$ 0.9 &  2.1$\pm$0.1 &  16.0$\pm$ 0.3 & 0.9$\pm$0.1 &    66 &   60 &  46.8 \\
SL\,435      & 0.6$\pm$0.1 &  9.0$\pm$ 2.8 & 22.6$\pm$13.9 &  3.3$\pm$0.1 &  23.6$\pm$ 0.6 & 0.9$\pm$0.1 &    57 &   54 &  73.4 \\
SL\,441      &   ...        &       ...      &      ...       &     ...       &        ...      &     ...      &    54 &   54 &  46.8 \\
SL\,472      &   ...        &       ...      &      ...       &     ...       &        ...      &     ...      &    60 &   51 & 100.5 \\
\hline
\multicolumn{10}{c}{LMC 8$\_$8} \\\hline
KMHK\,1510~  & 0.7$\pm$0.2 &  2.1$\pm$ 1.5 &  4.2$\pm$ 5.7 & 20.0$\pm$0.9 & 227.1$\pm$ 9.1 & 1.1$\pm$0.1 &   27 &   24 &   26.7 \\
BSDL\,3174   & 3.2$\pm$1.0 & 23.1$\pm$ 5.7 & 13.8$\pm$ 7.3 &  3.5$\pm$0.1 &  38.0$\pm$ 1.0 & 1.0$\pm$0.1 &   39 &   33 &   ...   \\
KMHK\,1516   & 4.6$\pm$1.5 & 34.6$\pm$ 7.6 & 16.7$\pm$ 8.0 &  5.5$\pm$0.3 &  84.0$\pm$ 4.2 & 1.2$\pm$0.1 &   72 &   72 &   60.3 \\
KMHK\,1519   & 1.0$\pm$0.1 &  9.1$\pm$ 2.5 & 13.9$\pm$ 6.2 & 18.0$\pm$0.4 & 181.0$\pm$ 4.0 & 1.0$\pm$0.1 &   39 &   39 &   46.7 \\
KMHK\,1521   & 2.1$\pm$0.2 & 43.1$\pm$ 5.2 & 38.4$\pm$ 8.2 & 19.0$\pm$0.3 & 200.1$\pm$ 2.9 & 1.0$\pm$0.1 &   72 &   66 &   53.3 \\
LW\,334      & 1.9$\pm$0.4 &  7.5$\pm$ 2.1 &  6.5$\pm$ 3.3 &  4.0$\pm$0.4 &  73.5$\pm$ 5.7 & 1.3$\pm$0.1 &   36 &   33 &   ...   \\
BSDL\,3188   & 1.2$\pm$0.4 &  8.0$\pm$ 3.8 &  9.1$\pm$ 8.2 & 19.0$\pm$0.6 & 212.2$\pm$ 6.6 & 1.1$\pm$0.1 &   51 &   42 &   ...   \\
KMHK\,1555   & 1.5$\pm$0.1 & 25.0$\pm$ 3.2 & 28.9$\pm$ 6.0 &~12.5$\pm$0.9 &~209.0$\pm$10.2 &~1.2$\pm$0.1 &   90 &   90 &   80.4 \\
KMHK\,1552   & 7.3$\pm$3.7 &~43.0$\pm$11.9 &~14.7$\pm$10.0 &  3.0$\pm$0.1 &  30.0$\pm$ 0.7 & 1.0$\pm$0.1 &   60 &   51 &   60.3 \\
KMHK\,1568   &        ...   &       ...        &       ...      &      ...      &         ...     &       ...    &   30 &   27 &   33.6 \\
KMHK\,1567   & 1.3$\pm$0.2 & 14.0$\pm$ 3 & 16.5$\pm$ 7.5 & 11.0$\pm$0.8 & 177.0$\pm$10.3 & 1.2$\pm$0.1 &   57 &   48 &   40.2 \\
KMHK\,1577   & 1.8$\pm$0.5 & 13.5$\pm$ 4.3 & 11.5$\pm$ 7.3 &  6.0$\pm$0.5 & 107.3$\pm$ 7.9 & 1.3$\pm$0.1 &   45 &   42 &   60.3 \\
KMHK\,1578   &        ...   &         ...    &         ...    &        ...    &          ...    &       ...    &   54 &   45 &   40.2 \\
KMHK\,1585   & 1.1$\pm$1.2 &  2.0$\pm$ 1.9 &  1.5$\pm$ 3.7 &  9.0$\pm$1.1 & 182.4$\pm$21.3 & 1.3$\pm$0.2 &   39 &   36 &   40.2 \\
KMHK\,1589   & 4.9$\pm$1.4 & 37.5$\pm$ 7.3 & 18.1$\pm$ 7.5 &  4.5$\pm$0.2 &  65.0$\pm$ 2.1 & 1.2$\pm$0.1 &   66 &   60 &   80.4 \\
KMHK\,1592   & 0.9$\pm$0.1 & 23.5$\pm$ 2.6 & 43.8$\pm$ 8.0 & 33.0$\pm$0.4 & 247.0$\pm$ 3.0 & 0.9$\pm$0.1 &~~132 &  114 & ~100.5 \\
KMHK\,1597   & 5.0$\pm$1.4 & 22.0$\pm$ 3.9 & 10.4$\pm$ 3.9 &  2.5$\pm$0.1 &  35.0$\pm$ 1.4 & 1.2$\pm$0.1 &   60 &   54 &   46.7 \\
KMHK\,1600   &        ...   &        ...     &         ...    &        ...    &          ...    &       ...    &   66 &   66 &   46.7 \\
KMHK\,1607   & 8.0$\pm$3.7 & 20.0$\pm$ 5.0 &  6.7$\pm$ 4.1 &  0.9$\pm$0.1 &  14.0$\pm$ 1.0 & 1.2$\pm$0.2 &   48 &   39 &   53.3 \\
KMHK\,1609   & 1.4$\pm$0.2 & 20.7$\pm$ 3.8 & 24.6$\pm$ 7.4 & 16.0$\pm$0.4 & 193.2$\pm$ 4.2 & 1.1$\pm$0.1 &   90 &   84 &   93.5 \\
KMHK\,1611   & 1.0$\pm$0.1 &  5.0$\pm$ 1.5 &  7.3$\pm$ 3.4 & 10.5$\pm$0.9 & 202.2$\pm$14.4 & 1.3$\pm$0.2 &   39 &   33 &   40.2 \\
KMHK\,1623   & 2.6$\pm$0.5 & 15.5$\pm$ 2.7 & 11.4$\pm$ 3.8 &  3.5$\pm$0.3 &  63.5$\pm$ 3.7 & 1.3$\pm$0.1 &   72 &   72 &   53.3 \\
\hline
\end{tabular}
\medskip

\normalsize

\noindent Note: The 1-$\sigma$ errors are listed.
For comparison we list the cluster sizes from
\citet{betal08} and \citet{ketal90}, in arcsec. The sizes of \citet{ketal90}
were converted from pc to arcsec units using a LMC distance modulus of
$\mu$=18.49\,mag.

\end{table*}

\clearpage

\begin{figure}
\centerline{\psfig{figure=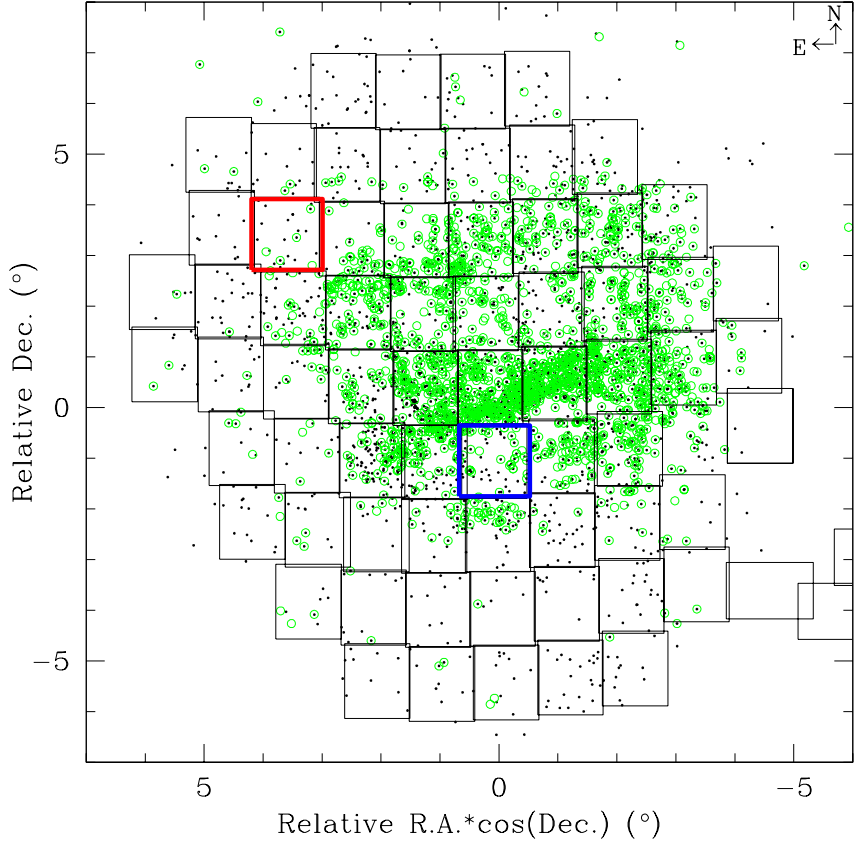,width=114mm}}
\caption{Sky-projected spatial distribution of B08's catalogue of star clusters in the LMC 
centred at $RA=05:23:34$, $DEC= -69:45:22$ (J2000). 
Black points and green circles represent catalogued star clusters 
and those with age estimates available, respectively. The objects studied in this work are placed
in tiles LMC 5$\_$5 (blue rectangle) and LMC 8$\_$8 (red rectangle).}
\label{fig1}
\end{figure}

\begin{figure}
\centerline{\psfig{figure=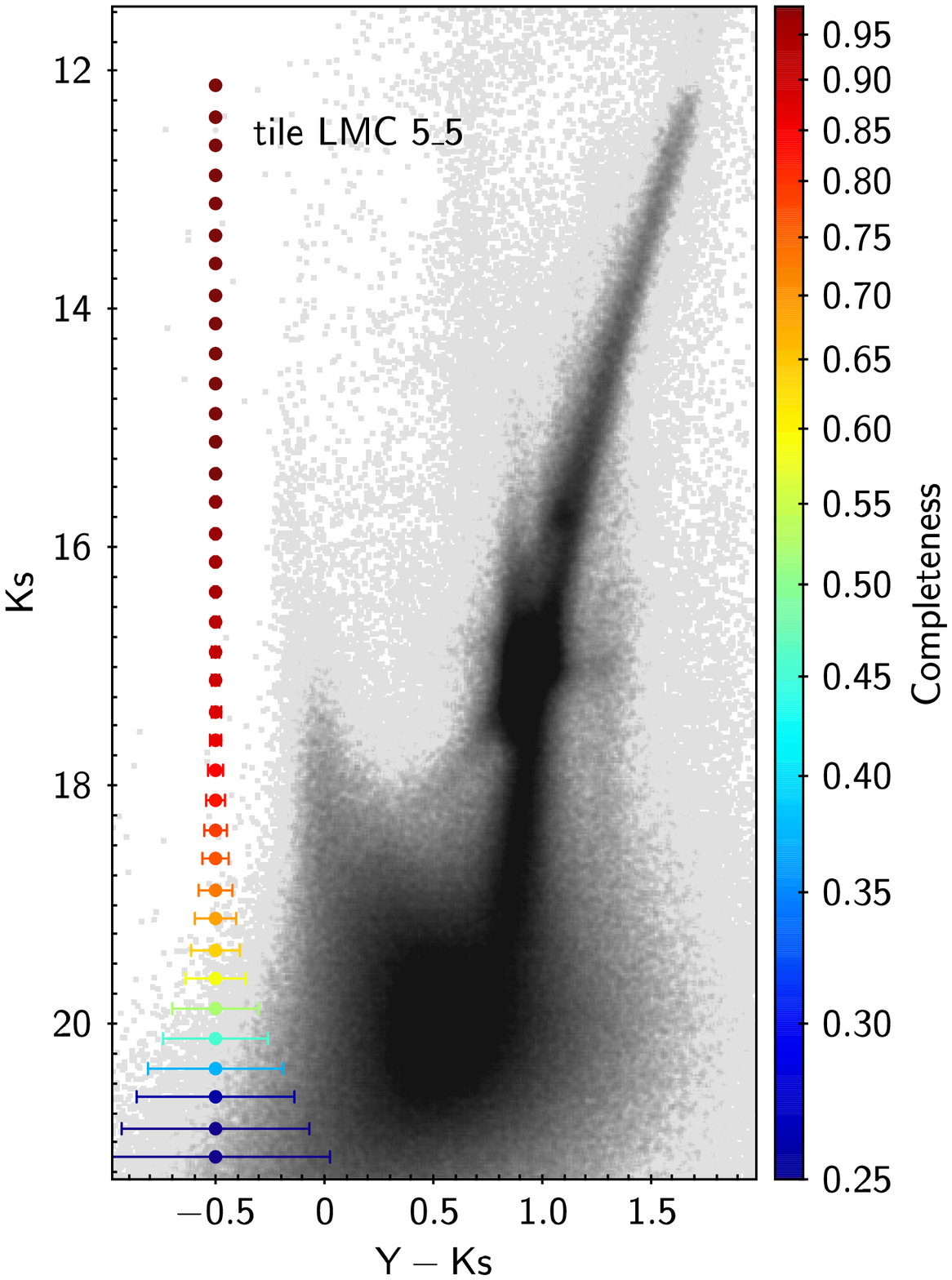,width=100mm}}
\centerline{\psfig{figure=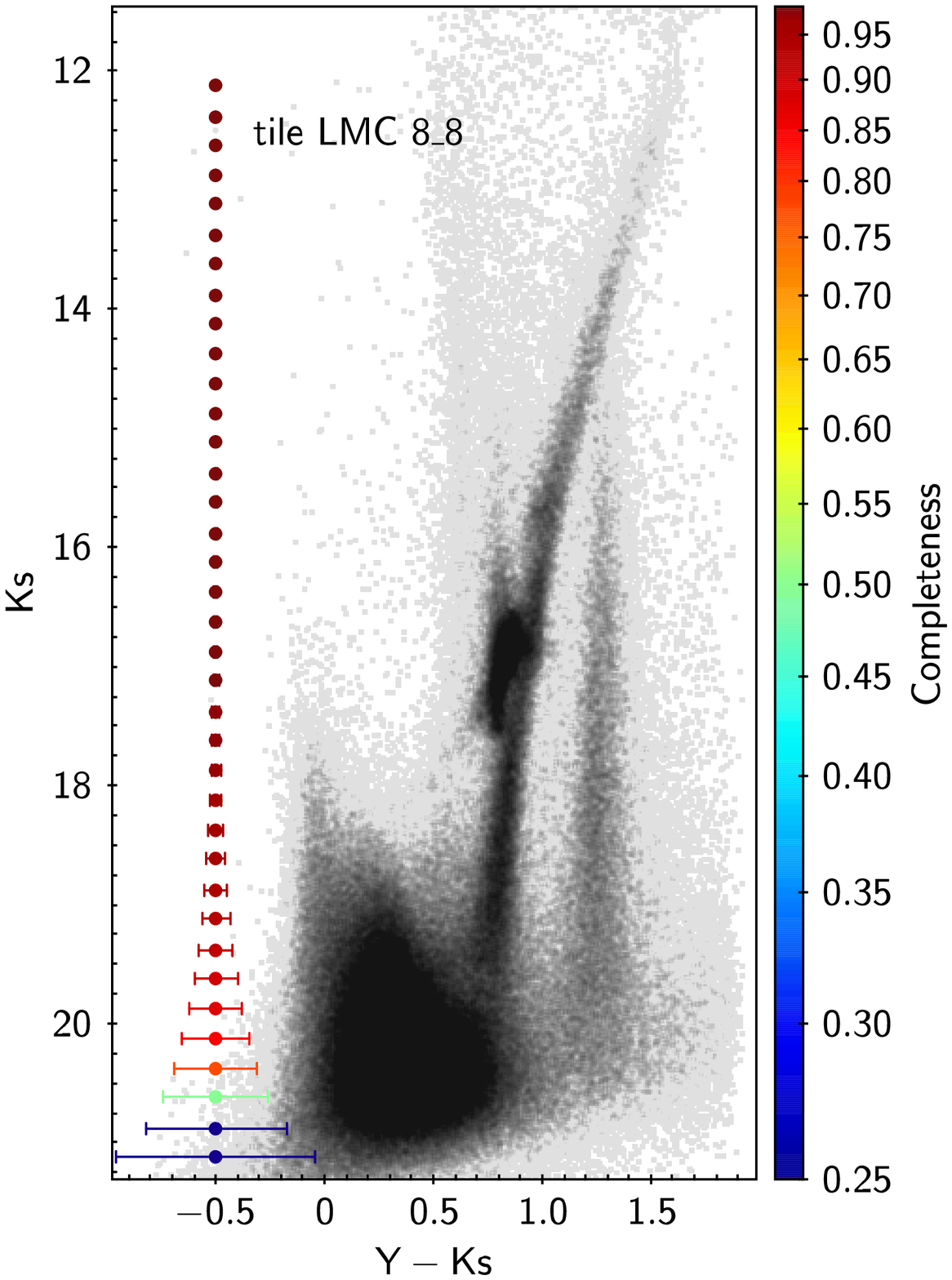,width=100mm}}
\caption{CMDs for stars in tiles LMC\,5\_5 and 8\_8 with error bars coloured according to the
colour scale of the completeness level.}
\label{fig2}
\end{figure}

\begin{figure}
\centerline{\psfig{figure=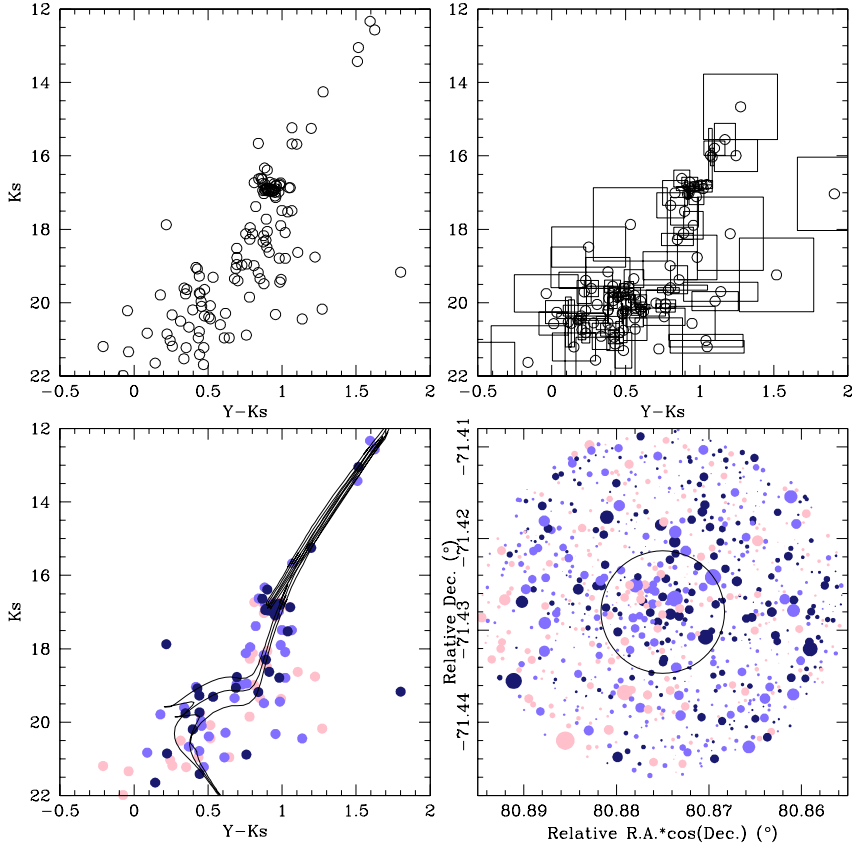,width=114mm}}
\caption{CMDs for stars in the field SL\,435 (LMC 5$\_$5): the observed CMD for the stars distributed 
within the cluster radius (upper-left panel); a field CMD for an annulus centred on the 
cluster, with an internal radius 3 times the cluster radius and an area equal to that of the cluster area
(upper-right panel); the cleaned cluster CMD (bottom-left). We overplotted designed box-shaped cells
for each star in the field CMD to be used in the cluster CMD field decontamination (see section 4 for details). 
Colour-scaled symbols represent 
stars that statistically belong to the field (P $\le$ 25\%, pink), stars that might 
belong either to the field or to the cluster (P  $=$ 50\%, light blue), and stars that 
predominantly populate the cluster region (P $\ge$ 75\%, dark blue). Three isochrones from Marigo et 
al. (2008) for log($t$/yr), log($t$/yr) $\pm$ 0.1, and metallicity values listed in Table 1
 are also superimposed. The 
schematic chart centred on the cluster for a circle of radius 3 times the cluster radius
is shown in the bottom-right panel. The black circle represents the adopted cluster radius. 
Symbols are as in the bottom-left panel and with a size proportional to 
the brightness of the star. North is upwards, and East is to the left.}
\label{fig3}
\end{figure}

\clearpage
\begin{figure}
\centerline{\psfig{figure=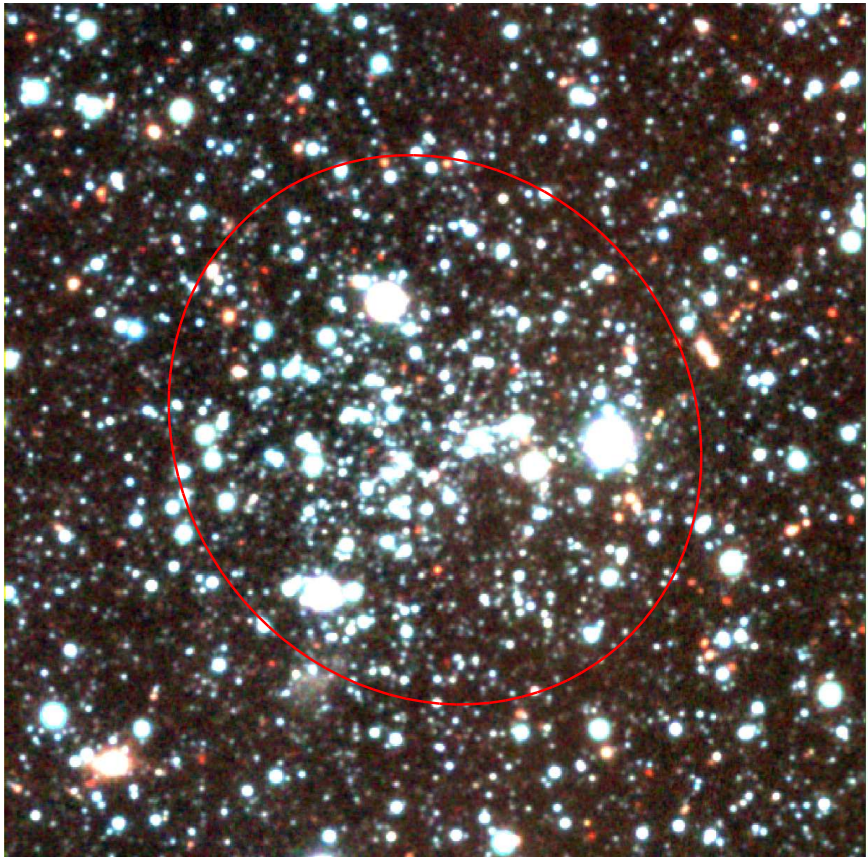,width=114mm}}
\centerline{\psfig{figure=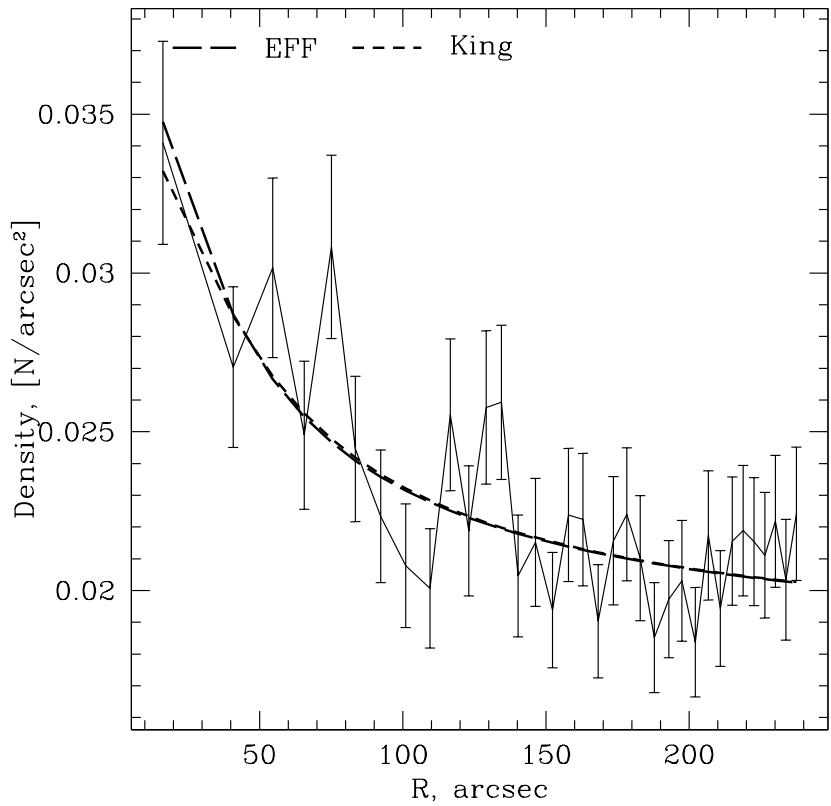,width=114mm}}
\caption{Top: Three-colour composite 3$\times$3\,arcmin image of KMHK\,1592 (LMC 8$\_$8)
($Y$ - blue, $J$ - green, $K_\mathrm{s}$ - red; North is up and East is to the left.). Bottom: profiles of KMHK\,1592
obtained from the King and EFF models compared with the data.}
\label{fig4}
\end{figure}

\clearpage
\begin{figure}
\centerline{\psfig{figure=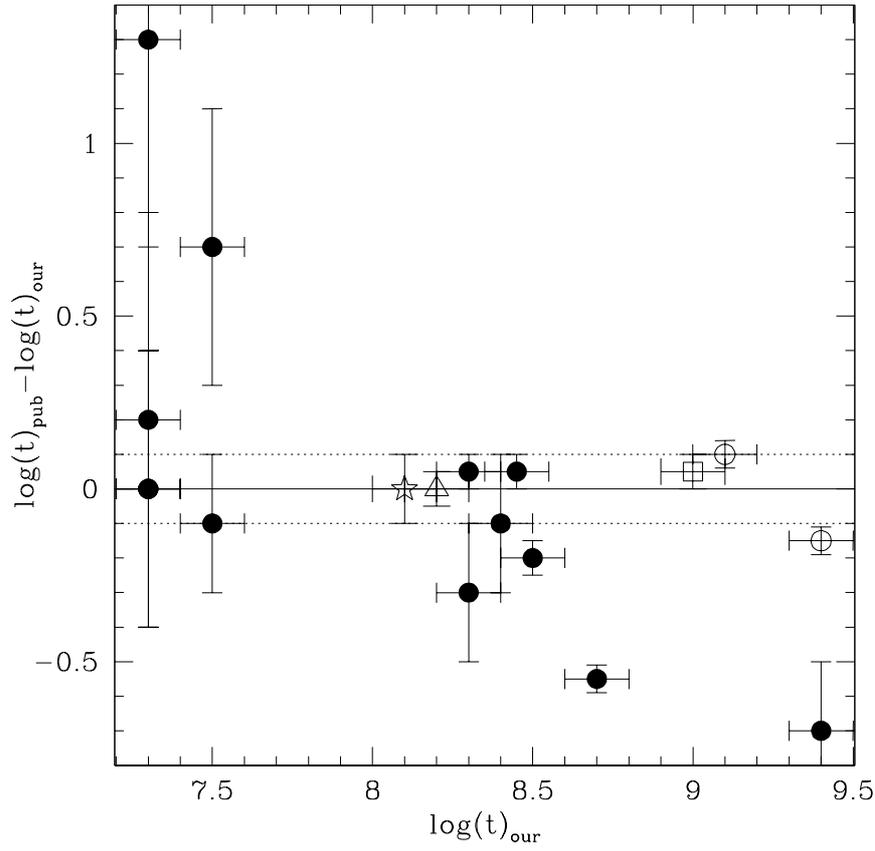,width=114mm}}
\caption{Age difference between present age estimates and those published by Glatt et al.
(2010, filled circle), Piatti (2011, open circle), Milone et al. (2009, open box),
Gouliermis et al. (2010, open triangle), and Piatti (2012, open star). The vertical
errorbars represent the age uncertanties from the published values.}
\label{fig5}
\end{figure}
\clearpage
\begin{figure}
\centerline{\psfig{figure=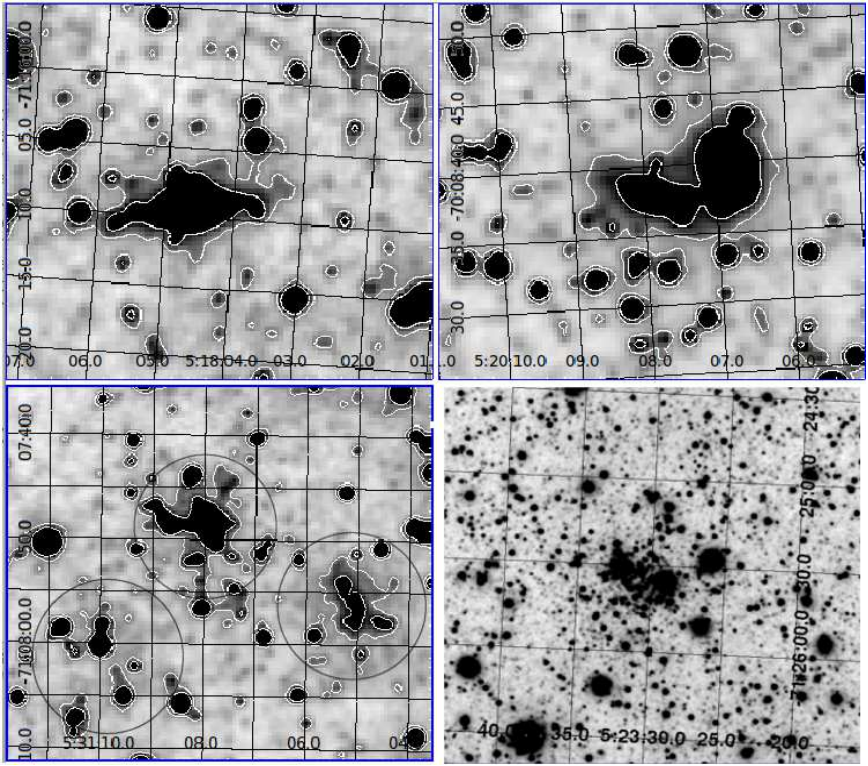,width=114mm}}
\caption{Enlargement of $K_s$ images of tile LMC 5$\_$5 centred on KMHK\,747 (upper-left, possible galaxy), 
OGLE\,366 (upper-right, possible galaxy), BSDL\,2144 (bottom-left, possible triple system), and
SL\,435 (bottom-right). Isophote curves have been superimposed.}
\label{fig6}
\end{figure}
\clearpage

\begin{figure}
\centerline{\psfig{figure=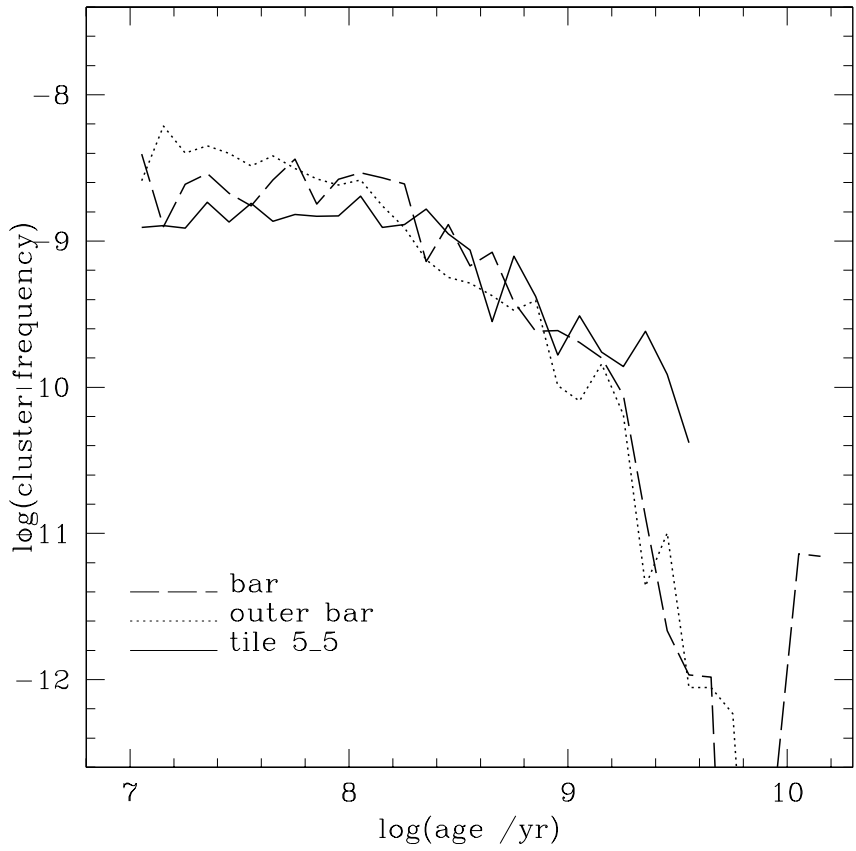,width=114mm}}
\caption{CFs for the LMC outer Bar, Bar and tile LMC 5$\_$5 (see text for details).}
\label{fig7}
\end{figure}

\end{document}